\documentclass[pre,aps,twocolumn]{revtex4}

\usepackage{amsmath,amssymb}
\usepackage{graphicx}

\begin{document}

{\bf Comment on "Liquids on Topologically Nanopatterned Surfaces"}\\

In a recent letter, Gang {\it et al.} report measurements of liquid adsorption on substrates with geometrical structure on the nanometric scale \cite{Pershan}. This study is particularly interesting for a number of reasons: the chosen geometry (paraboloidal cavities), the size of the structure (in the nanometric range) and the use of two concording experimental methods (x-ray reflectivity and grazing incident diffraction). In the paper, comparison is made with the predictions of a very simple "geometrical" model for adsorption on sculpted substrates \cite{Nat}. The authors compare their results with an estimation of the (asymptotic) power-law prediction of the geometrical model and conclude that they are significantly different. Here we point out that full application of the geometrical model for a finite-size (FS) paraboloid yields results which compare favourably with their experimental findings. This is to a certain extent surprising, due to the small scale of the structures, and supports previous evidence of the strong influence of surface geometry on fluid adsorption \cite{Mistura}.

\begin{figure}[thb]
\hspace*{-.5cm}
\includegraphics[width=7.5cm]{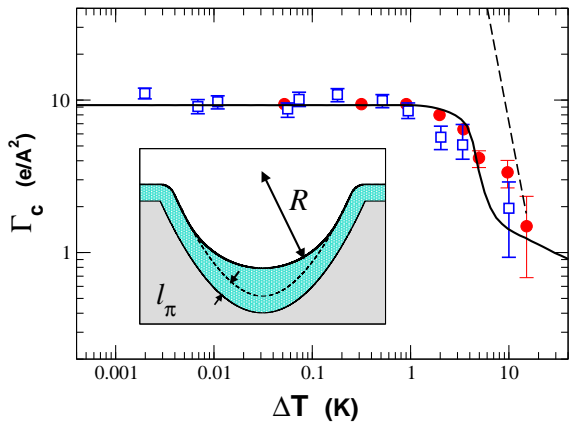}
\caption{Liquid adsorption $\Gamma_c$ in a finite paraboloid: experimental points (symbols), prediction of the geometrical model (continuous line) and power-law estimate as published by Gang {\it et al.\ }\cite{Pershan} (dashed line). This line only represents the slope of the power-law, not its absolute value \cite{Pershan2}} \label{fig1}
\end{figure}

The inset in Fig.\ 1 schematically illustrates the geometrical construction as applied to a FS paraboloid. The liquid gas-interface is obtained by first coating the substrate with a layer of thickness $\ell_\pi$ followed by fitting a meniscus of radius $R$ at the point of maximum curvature \cite{Nat}. This construction requires only two length scales, the thickness of the liquid layer adsorbed on a \textit{flat} substrate $\ell_\pi$ and the radius of curvature given by the Laplace equation $R$ \cite{Nat}. Both these quantities depend on the chemical potential $\Delta\mu$, relative to liquid-gas coexistence. Indeed, for our particular case, we have\vspace*{-.15cm}
\begin{eqnarray}
\ell_\pi(\Delta\mu)=\left(\frac{2A}{\Delta\mu\,\Delta\rho}\right)^{1/3} \hspace*{.85cm} R(\Delta\mu)=\frac{2\sigma}{\Delta\mu\,\Delta\rho}
\end{eqnarray}
where $A=1.34\times 10^{-21}$ J is the Hamaker constant, $\sigma=23.42$ mN/m the liquid-gas surface tension and $\Delta\rho=4.6956$ nm$^{-3}$ the density difference between the coexisting phases \cite{Pershan,MCH}.

This procedure allows one to predict a number of geometrical quantities as a function of $\Delta\mu$, including the adsorption in the paraboloidal cavity, $\Gamma_c$. This quantity is plotted in Fig.\ 1 together with the experimental results of Gang {\it et al.} as a function of the temperature difference between the substrate and the gas $\Delta T$ (instead of $\Delta\mu$) in line with the authors \cite{Pershan}. Despite the simplicity of the model, there is an overall agreement between theory and experimental data. We want to emphasise here that the theory has {\it no adjustable parameters}. 

There are three regimes: I) For $\Delta T\gtrsim 8 K$, no meniscus is present and the adsorption is essentially $\ell_\pi A_r$, where $A_r$ is the {\it real} area of the substrate (as opposed to the projected area), II) For $5\lesssim\Delta T\lesssim 8 K$, the adsorption is strongly sensitive to saturation conditions due to the sudden rise of the liquid meniscus, and III) For $\Delta T\lesssim 5 K$, the meniscus is essentially "pinned" to the rim of the paraboloid and the increase in adsorption is only due to its changing radius $R$. As predicted \cite{Nat}, the rise of the meniscus is so abrupt in regime II that the finite paraboloid fills almost completely for a small change in $\Delta T$. Thus, the asymptotic regime is essentially undetectable (bound closely by regimes I and III) and a comparison of the experimental adsorption with a mere power-law (see Fig.\ \ref{fig1}) indicates unwarrantedly that the predictions of the geometrical model are inadequate for the finite paraboloid.

Note as well that the abrupt filling takes place at a value of $\Delta\mu$ strongly dependent on the geometry of the cavities and, therefore, any dispersity in the shape or size of the experimental cavities (apparent in Fig.\ 1(a) of \cite{Pershan}) will smooth the adsorption curve yielding a smaller (effective) value of the exponent $\beta_c$ and, perhaps, be responsible for the small discrepancies at high $\Delta T$.

\vspace*{.25cm}

\noindent C.\ Rasc\'{o}n\\[.15cm]
GISC, Departamento de Matem\'{a}ticas\\
Universidad Carlos III de Madrid, 28911 Legan\'{e}s, Spain.\\[-.15cm]


\noindent Acknowledgements: Mossnoho and Mosaico grants.


\begin{thebibliography}{99}
\bibitem{Pershan} O.\ Gang {\it et al}, Phys.\ Rev.\ Lett.\ {\bf 95}, 217801 (2005).
\bibitem{Nat} C.\ Rasc\'{o}n, A.O.\ Parry, Nature {\bf 407}, 986 (2000).
\bibitem{Mistura} L.\ Bruschi {\it et al}, Phys.\ Rev.\ Lett.\ {\bf 89} 166101 (2002).
\bibitem{MCH} A.\ Villares {\it et al}, J.\ Solution Chem.\ {\bf 34}, 185 (2005).
\bibitem{Pershan2} O.\ Gang {\it et al}, (erratum).
\end{thebibliography}
\end{document}